# *In situ* observation of thermally activated and localized Li leaching from lithiated graphite

Harrison Szeto[1], Vijay Kumar[2], Yangying Zhu[2,*]

[1.] Department of Chemistry and Biochemistry, University of California Santa Barbara, Santa Barbara, CA 93106, USA
[2.] Department of Mechanical Engineering, University of California Santa Barbara, Santa Barbara CA 93106, USA

Email: Yangying Zhu (yangying@ucsb.edu)

**Abstract**

Temperature is known to impact Li-ion battery performance and safety, however, understanding its effect on Li-ion batteries has largely been limited to uniform high or low temperatures. While the insights gathered from such research are important, much less information is available on the effects of non-uniform temperatures which more accurately reflect the environments that Li-ion batteries are exposed to in real world applications. In this paper, we characterize the impact of a microscale, temperature hotspot on a Li-ion battery using a combination of *in situ* micro-Raman spectroscopy, *in situ* optical microscopy and COMSOL Multiphysics thermal simulations. Our results show that mild temperature heterogeneity induced by the micro-Raman laser can cause lithium to locally leach out from different lithiated graphite phases ($LiC_6$ and $LiC_{12}$) in the absence of an applied current. The Li metal is found to be largely localized to the region heated by the micro-Raman laser and is not observed upon uniform heating to comparable temperatures suggesting that temperature heterogeneity is uniquely responsible for causing Li to leach out from lithiated graphite phases. A mechanism whereby localized temperature heterogeneity induced by the laser induces heterogeneity in the degree of lithiation across the graphite anode is proposed to explain the localized Li leaching. This study highlights the sensitivity of lithiated graphite phases to minor temperature heterogeneity in the absence of an applied current.

**TOC Graphic**

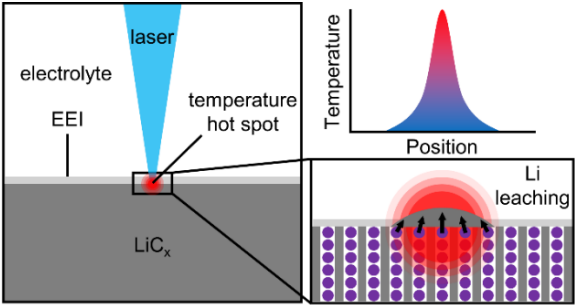

The ability of Li-ion batteries to provide cheap and modular energy has made them an important form of energy storage.[1–3] As Li-ion batteries become more broadly adopted across industries, there remains a need to understand and address sources of battery degradation to ensure their optimal performance and safe operation.[4–6] One source of degradation known to negatively impact Li-ion battery performance is temperature with high temperatures generally understood to lead to accelerated degradation via promotion of electrode-electrolyte interphase (EEI) growth at the anode surface as well as transition metal dissolution and oxygen evolution at the cathode surface.[7–11] Whereas the mechanisms surrounding degradation of Li-ion batteries at high temperatures are primarily thermodynamically-driven, low temperatures are associated with sluggish kinetics as a result of higher charge transfer resistances as well as lower Li diffusion at both electrodes.[7,12,13] These processes collectively result in loss of Li inventory as well as loss of active material which ultimately reduce battery capacity, cycling rate performance, and can even lead to thermal runaway.[7–14] While the aforementioned insights have been drawn from studies seeking to understand the effects that homogeneous, low or high temperatures have on Li-ion battery performance, there is a much more limited body of work exploring the effects of non-uniform temperatures.[15,16] Thus, there is considerable need to investigate the effects of non-uniform temperatures on Li-ion batteries to provide a more nuanced understanding of how they are impacted by temperature.

In recent years, a few preliminary studies have attempted to examine the effects of non-uniform temperatures on Li-ion batteries and reported unique degradation phenomena occurring as a result of temperature heterogeneity.[17–22] Initial studies examining the effects of a temperature hotspot on Li electrodeposition have reported that Li preferentially deposited in the region of the temperature hotspot due to locally enhanced surface exchange current density.[18,19] Similar results were reported when inducing a temperature hotspot on the surface of a graphite anode during lithiation which resulted in Li plating localized to the temperature hotspot, however, in this study the localized Li plating was attributed to a local increase in the temperature-dependent equilibrium Li plating potential above the equilibrium potential of graphite as it was lithiated.[20] Other work exploring the effects of an interelectrode thermal gradient on battery performance found that mild temperature gradients between an anode and cathode induced rapid capacity degradation with the degradation mode experienced by the cell determined by the direction of the temperature gradient.[17,21,22] While these studies clearly illustrate the effects that temperature heterogeneity have on Li-ion batteries, the non-uniform temperature conditions were all applied to batteries during operation and thus the degradation modes induced are electrochemical in nature. An *in situ* study would therefore provide complementary information on the thermal and/or chemical stability of a battery in response to temperature heterogeneity.

In this paper, we report thermally activated and localized lithium leaching from lithiated graphite phases ($LiC_6$ and $LiC_{12}$) in response to mild temperature heterogeneity. Lithium leaching is a phenomenon where lithium has been found to deintercalate from lithiated graphite phases under elevated, isothermal temperatures in the absence of an applied current.[23–28] In this work lithium leaching is observed by utilizing *in situ* micro-Raman spectroscopy, optical microscopy, and COMSOL simulations. Micro-Raman spectroscopy is employed to simultaneously induce a tunable laser hotspot on a single graphite particle embedded within a graphite electrode while also serving as a method for tracking the occurrence of lithium leaching. The laser hotspot represents a non-uniform temperature distribution which a Li-ion battery may be exposed to. *In situ* Raman spectra taken at the surface of lithiated graphite species reveal the appearance of a $Li_2C_2$ peak localized to the temperature hotspot which is associated with the EEI of Li metal.[29–31] The occurrence of lithium leaching is shown to be dependent on the laser power density with higher laser power densities inducing larger temperature variation across the electrode and more severe localized lithium leaching from $LiC_6$ and $LiC_{12}$. A mechanism whereby localized temperature heterogeneity induced by the laser induces heterogeneity in the degree of lithiation across the graphite anode is proposed to explain the localized lithium leaching from lithiated graphite phases.

In order to probe the effect of a temperature hotspot on graphite, a custom cell was designed as shown in Figure S1 which was modified with an optical window as well as holes in the layers above the graphite electrode to provide optical access for *in situ* measurements. The *in situ* cell is based on a modified half-cell architecture with a Li metal counter electrode and graphite working electrode. Additional details on the *in situ* cell construction and its electrochemical performance can be found in the Supporting Information.

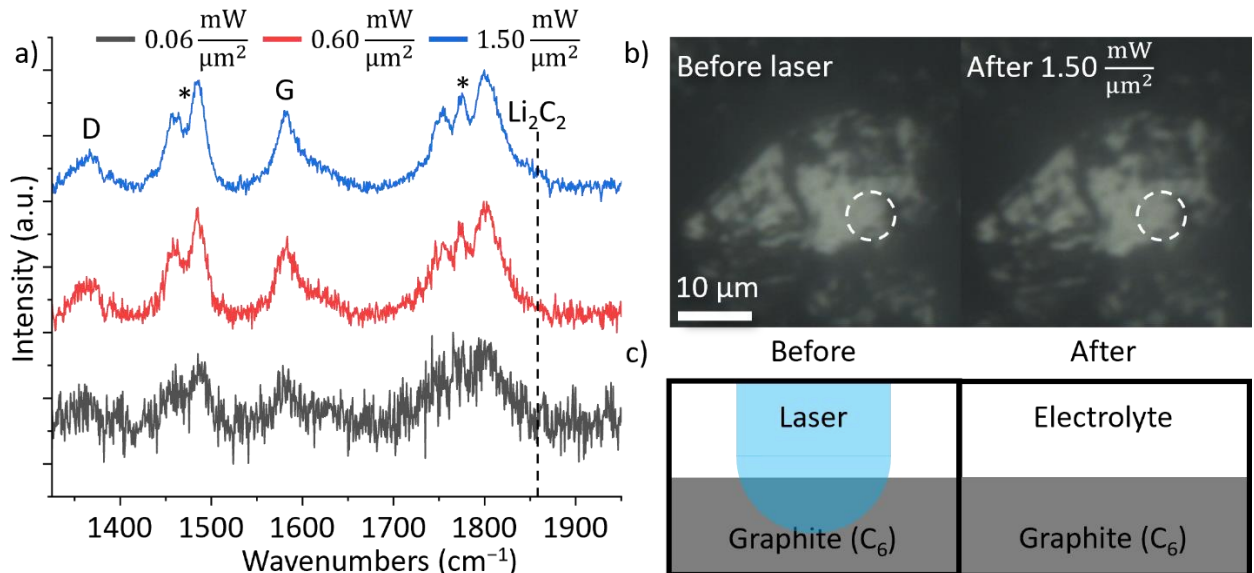

**Figure 1:** (a) Raman spectra for a graphite anode particle acquired before lithiation with different laser power densities. * denotes Raman bands from the electrolyte which is composed of 1M $LiPF_6$ in 1:1 by vol. EC:DMC. Additional peaks include graphite D and G bands. (b) Optical microscopy images taken before and after 1.50 mW/μm² laser exposure in the unlithiated state. The dashed white circle indicates the region where the laser irradiated the graphite particle. (c) Schematic depicting the no change on the surface of unlithiated graphite upon exposure to micro-Raman laser.

Initial Raman spectra taken on the surface of a graphite particle embedded within a graphite electrode before lithiation (Figure 1a) consists of the expected D and G bands of graphite at ~1350 cm$^{-1}$ and 1580 cm$^{-1}$ as well as several, prominent bands associated with the electrolyte (1 M LiPF$_6$ in 1:1 by vol. EC:DMC) denoted by an *.[32–34] A more detailed assignment of the electrolyte Raman bands can be found in Table S2. *In situ* optical microscopy images of the graphite particle taken before laser exposure and after the highest laser power density (1.50 mW/μm$^2$) are shown in Figure 1b with the white, dashed circle indicating the region exposed to the Raman laser during spectrum acquisition. A more complete series of optical microscopy images taken after each laser power density is provided in Figure S3. As expected, no changes are observed on the graphite particle before and after laser exposure which in addition to the Raman spectra, suggests the surface is thermally stable in its current environment and under the Raman acquisition parameters utilized as indicated in Figure 1c.

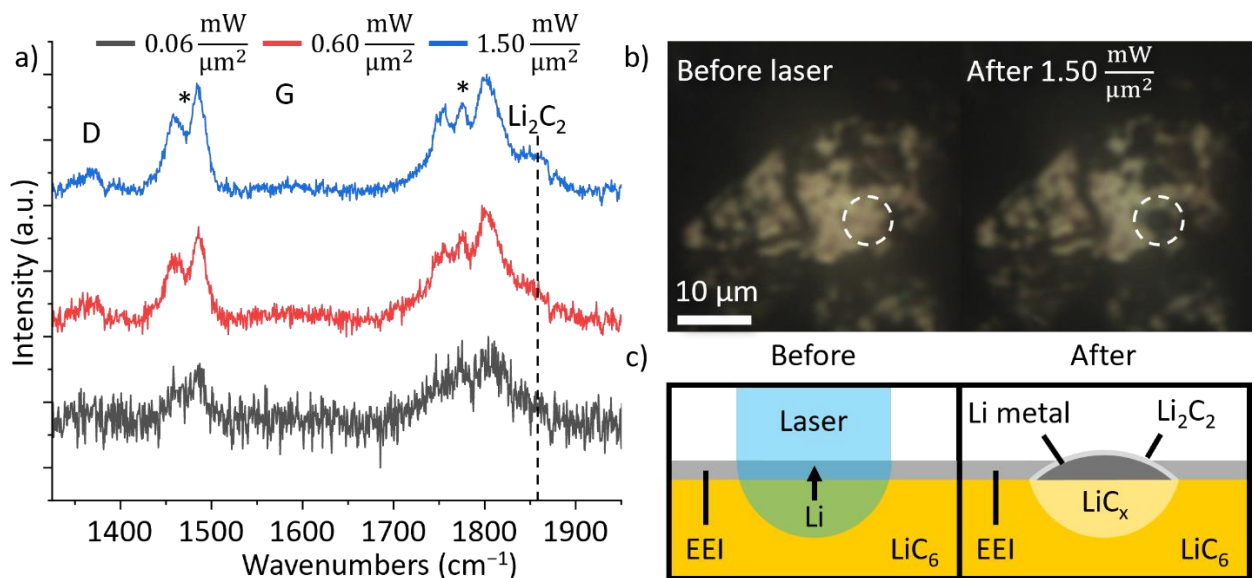

**Figure 2:** (a) Raman spectra for fully lithiated graphite (LiC$_6$) particle acquired with different laser power densities. An additional peak ~1850 cm$^{-1}$ appears which is attributed to Li$_2$C$_2$ which is indicative of lithium leaching. (b) Optical microscopy images taken before and after laser exposure in the lithiated state. Note that the portion of the sample irradiated by the laser becomes slightly

discolored after exposure. (c) Schematic depicting lithium leaching from lithiated graphite upon exposure to the micro-Raman laser.

After lithiation of the graphite to $LiC_6$ under a rate of C/10 at room temperature, *in situ* micro-Raman spectroscopy is employed to examine $LiC_6$ (Figure 2a) where the following changes are observed: (1) the disappearance of the D and G bands and (2) the appearance of a new peak ~1850 cm$^{-1}$. While the disappearance of the D and G bands is expected following the lithiation of graphite,[35,36] the peak ~1850 cm$^{-1}$ is attributed to $Li_2C_2$ which has previously been found on the surface of Li metal as well as on Li plated on graphite under fast charging or overcharging conditions.[29–31,37–41] Importantly, however, the lithiation conditions utilized in this work are mild and therefore not conducive with Li plating on the surface of $LiC_6$.[42,43] Interestingly, the appearance of the $Li_2C_2$ coincides with a change in color on the surface of $LiC_6$ that is localized to the region exposed to the laser (Figure 2b). As the laser power density is increased, there is an increase in the intensity of the $Li_2C_2$ peak relative to the other electrolyte peaks that occurs concurrently with an increase in the extent of discoloration on the surface of $LiC_6$ (Figure S4). These results suggest that the laser is locally heating $LiC_6$ which promotes lithium to leach out of $LiC_6$ and subsequently leads to an increase in the abundance of $Li_2C_2$ as represented in Figure 2c.

In order to further elucidate the origin of $Li_2C_2$, an adjacent region on the same graphite particle was examined in the delithiated state (Figure S5) which revealed the expected reappearance of the graphite D and G bands and importantly the absence of the $Li_2C_2$ peak at ~1850 cm$^{-1}$. The cell was delithiated prior to spectrum acquisition due to the sensitivity of $LiC_6$ to even relatively low laser power densities. While it is not expected that the lithiation conditions utilized would lead to Li plating, if Li were to have plated on the graphite, active Li would be stripped during delithiation but the $Li_2C_2$ would be expected to persist on the delithiated graphite. Thus, the absence of a peak at ~1850 cm$^{-1}$ on a region adjacent to the spot exposed to the laser in the LiC$_6$ state confirms that Li$_2$C$_2$ is localized and its formation is thermally activated. These results are also consistent with optical microscopy images taken after *in situ* Raman spectra acquisition in each of the aforementioned states whereby a localized discoloration of the region exposed to the laser occurs in the fully lithiated state (LiC$_6$) but does not occur in the unlithiated and delithiated states, even at higher laser power densities (Figures S2, 3, 6).

Subsequent experiments carried out to probe the thermal sensitivity at the intermediate stage II (LiC$_{12}$) reveal largely similar results. As shown in Figure S7a, Raman spectra for LiC$_{12}$ largely resemble those for LiC$_6$ with electrolyte peaks as well as the graphite D band, a slightly more visible G band, and Li$_2$C$_2$. The Li$_2$C$_2$ is observed at a comparable laser power density to LiC$_6$ with localized discoloration also observed, however, for higher power densities, there is a dramatic change in the whole color of the graphite particle from red to gold (Figure S7b). Optical microscopy images taken after each laser power density are also provided in Figure S8 to show how the discoloration evolves. Interestingly, the dramatic color change coincides with a decrease in the intensity of the graphite G band which collectively suggests that there may have been a shift in phase equilibrium towards LiC$_6$ on the surface of the particle.

The findings obtained from *in situ* micro-Raman spectroscopy and optical microscopy images suggest that both LiC$_6$ and LiC$_{12}$ exhibit a temperature sensitivity whose distribution is of interest. While development of an experimental methodology to measure the temperature induced by the laser hotspot on the graphite surface is preferable, such an endeavor requires significant development of a non-invasive, nano-thermometry method that is electrochemically compatible with the graphite particle, which is non-trivial in and of itself and thus beyond the scope of this study. Given the challenges associated with making such a measurement, the Heat Transfer in Solids module of COMSOL Multiphysics is instead employed to estimate the temperature distribution induced by the laser hotspot on graphite and its other lithiated phases. As shown in Figure 3a, the 2D-axisymmetric model consists of a graphite particle embedded within a porous, composite graphite electrode and surrounded above by electrolyte. Additional information on the model can be found in the Supporting Information.

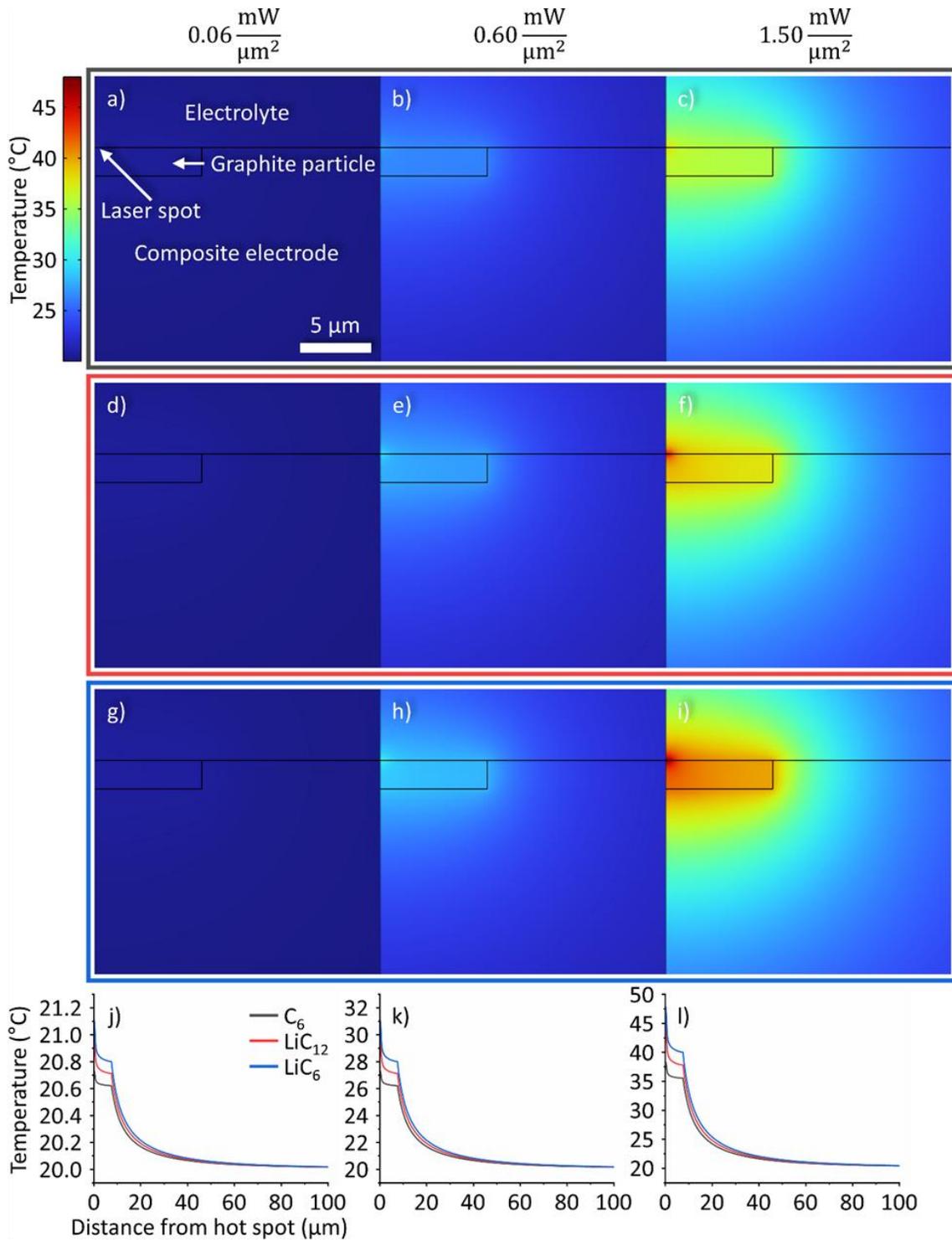

**Figure 3:** 2D temperature distribution generated in COMSOL for different phases of lithiated graphite and different laser power densities. The 2D temperature distribution induced on $C_6$ for different laser power densities are shown in (a), (b), and (c). The 2D temperature distribution induced on $LiC_{12}$ for different laser power densities are shown in (d), (e), and (f). The 2D temperature distribution induced on $LiC_6$ for different laser power densities are shown in (g), (h),

and (i). 1D temperature plot generated for the different phases of lithiated graphite at (j) 0.06 mW/μm$^2$, (k) 0.60 mW/μm$^2$, and (l) 1.50 mW/μm$^2$.

The temperature distribution induced by the laser hotspot on pristine graphite, LiC$_{12}$, and LiC$_6$ for different power densities is shown in Figure 3 which confirms that the laser is inducing microscale temperature heterogeneity. The highest temperatures at each power density are largely confined to the region of the laser hotspot due to the poor thermal conductivities of the electrolyte (~0.1 W/mK) and bulk graphite electrode (~1 W/mK) surrounding the graphite particle. As the laser power density increases, there is a corresponding increase in the maximum temperature for all phases of lithiated graphite from ~20.7–21.0 °C for a laser power density of 0.06 mW/μm$^2$, to ~27.3–31.2 °C for 0.60 mW/μm$^2$ and ultimately ~38.4–48.0 °C for 1.50 mW/μm$^2$. In all of the simulations, the lowest temperature on the graphite electrode remains at room temperature (20.0 °C). In addition, the variation in the maximum temperature which increases with laser power density is due to the variation in the thermal conductivities of the different phases of lithiated graphite with Li intercalants reducing the thermal conductivity of LiC$_{12}$ and LiC$_6$ relative to C$_6$. These results coupled with optical microscopy images and Raman spectra further reinforce that the temperature hotspot is inducing localized lithium leaching. Furthermore, essentially all of the maximum temperatures induced by the laser are conventionally understood to have marginal impacts on battery safety and performance which represents a significant discrepancy between our findings and conventional understanding of the thermal sensitivity of LiC$_{12}$ and LiC$_6$.[23,24,26,27,44]

In order to determine whether the localized lithium leaching is occurring as the result of heating above a threshold temperature, additional experiments were performed involving uniform heating whereby graphite was fully lithiated under the same conditions as the samples described above (C-rate = C/10 at 20 °C), before being placed in an environmental chamber held at 40 °C for 1 hour. Given that our experiments have suggested LiC$_{12}$ and LiC$_6$ are prone to lithium leaching upon exposure at relatively low laser power densities, the cell was subsequently cooled to room temperature and delithiated at C/10 before being examined with micro-Raman spectroscopy. As shown in Figure S9, there is no $Li_2C_2$ peak visible in spectra taken for delithiated graphite (after uniform heating) even under very high laser power densities. In addition, optical microscopy images taken in the delithiated state (after uniform heating) show no visual discoloration which further confirms the absence of lithium leaching following uniform heating. These results suggest that temperature heterogeneity rather than a threshold temperature is responsible for inducing lithium leaching.

Considering the non-uniform temperature distribution induced by the micro-Raman laser estimated in COMSOL, it is important to consider the properties which are impacted by temperature. One such property is the electrochemical potential whose temperature dependence is denoted by the temperature coefficient (α) for a given half cell reaction potential as indicated by (1) and (2).

$$A + ne^- \rightarrow B \tag{1}$$

$$\alpha = \frac{\partial E_{eq}}{\partial T} \tag{2}$$

Given that α is with respect to a half cell reaction, the equilibrium electrochemical potential would ideally be measured with respect to a reference electrode, however, the reference electrode itself has its own temperature coefficient which prevents direct determination of α under isothermal conditions. Therefore, the electrochemical potential of a symmetric cell is typically measured under non-isothermal conditions where the temperature at one electrode is varied while the other is held at a constant temperature. Under the assumption that there are negligible contributions from the Seebeck effect (i.e., thermoelectric effect due to electron transport within the electrode) and Soret effect (i.e., ion transport within the electrolyte) (3), α can be approximated from the slope of a plot of the measured electrochemical potential at different temperatures.

$$\frac{\partial E_{Gr}}{\partial T} = \frac{\partial E_{eq}}{\partial T} + \frac{\partial E_{Seebeck}}{\partial T} + \frac{\partial E_{Soret}}{\partial T}; \frac{\partial E_{Seebeck}}{\partial T} + \frac{\partial E_{Soret}}{\partial T} \ll \frac{\partial E_{eq}}{\partial T} \quad (3)$$

It is important to clarify that the Seebeck effect here refers to the potential induced by a temperature gradient difference within a single electrode due to electron transport, whereas some work has defined the Seebeck effect in an electrochemical cell differently.[45,46]

Previous studies have estimated the temperature coefficient for a graphite electrode at different states of charge to be approximately 1 mV/K.[20,47] The positive shift is illustrated in Figure 4a which shows the potential profile of graphite at different temperatures assuming a constant temperature coefficient of 1 mV/K. For uniform heating there is a vertical shift from point A to B in Figure 4a where the potential of lithiated graphite increases but the degree of lithiation remains the same across the graphite electrode. In contrast, for a non-uniform temperature distribution on an electrode, regions with a relatively high temperature would be expected to have a higher potential (such as shifting from point A to point B locally) than the relatively low temperature regions if lithium were to remain uniformly distributed across the electrode. However, the electrode is presumed to be electronically well-connected and the Seebeck (~20 μV/K)[48,49] and Soret[46] effects are relatively small, thus the spatial variation in the potential across the electrode is believed to be minimal. It is therefore proposed that lithium must redistribute within the graphite electrode such that the potential is uniform across the electrode and the total amount of intercalated lithium is conserved. An example involving a linear temperature gradient varying between 20 and 40 °C on a graphite electrode is shown in Figure 4b. It is proposed that the lithium redistributes within graphite such that the potential at different temperatures becomes the same across the electrode. This is indicated in Figure 4a with arrows from B to C (the colder region) and D (the warmer region). For a microscale temperature hotspot, however, the vast majority of the electrode remains at room temperature and thus the electrode potential is presumed to be essentially unchanged. Meanwhile, the region of the temperature hotspot will increase to higher degrees of lithiation in order to maintain at the same potential with the rest of the electrode as shown with an arrow from E to F in Figure 4c/d. Depending on the initial degree of lithiation and temperature heterogeneity ($\Delta T$), there appears to be a threshold temperature variation such that the degree of lithiation in the warm region is greater than 1 (x > 1 for $Li_xC_6$) leading to lithium leaching. It is believed that lithium redistribution could occur for other electrode materials exposed to temperature heterogeneity which highlights the need to more rigorously characterize the thermophysical properties of battery electrodes.[50]

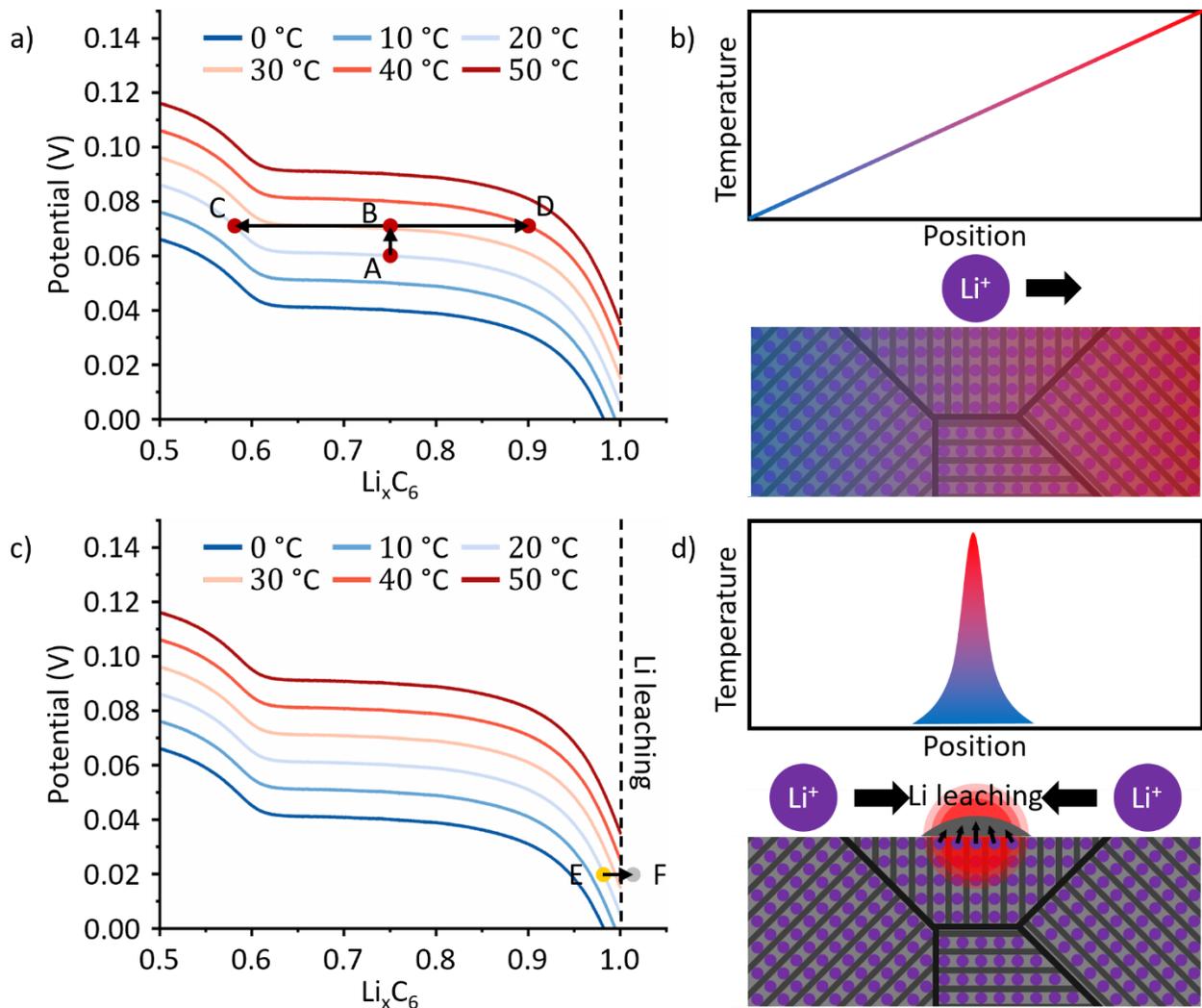

**Figure 4:** (a) Schematic plot of graphite electrode's potential response to uniform temperature and linear temperature gradient conditions. (b) Schematic showing lithium redistribution towards warmer region in graphite electrode with a linear temperature gradient. (c) Schematic plot of graphite electrode's potential response to localized temperature hotspot. (d) Schematic showing Li redistribution towards temperature hotspot and subsequent lithium leaching.

The implications of this study are important as they illustrate the importance of understanding the consequence of temperature heterogeneity on Li-ion batteries as well as the sensitivity of lithiated graphite phases at higher degrees of lithiation. While previous studies identifying how temperature heterogeneity during operation can result in non-uniform aging/lithiation, this work suggests additional consideration be given to temperature heterogeneity within a resting battery which may induce a non-uniform distribution of lithium within graphite.

Micro-Raman spectroscopy serves as a useful technique to facilely probe the effect of temperature hotspots in Li-ion batteries by tuning the micro-Raman laser power density. This study also highlights the care which must be taken when selecting acquisition parameters for micro-Raman spectroscopy and other characterization techniques involving lasers to analyze Li-ion batteries or battery materials. The poor thermal conductivity of the electrolyte and composite electrodes (~0.1 W/mK) coupled with the small irradiation area can lead to localized laser heating which may induce localized and thermally activated degradation phenomena similar to those reported in this paper. Further investigation may also be warranted for attempting to understand whether such degradation occurs when employing lasers in battery manufacturing where they have been used to cut and structure electrodes.[51–54]

While previously considered to be safe, the results of our study highlight the effects that exposure to heterogeneous temperatures have on a resting battery. These findings are particularly relevant for battery preconditioning where the battery thermal management system (BTMS) heats or cools the batteries to a more moderate temperature prior to charge or discharge.[55,56] The external nature of BTMSs makes them prone to inducing heterogeneous temperatures as the poor thermal conductivities of the electrolyte and composite electrodes (~0.1–1 W/mK) limit the efficiency of heat conduction to or away from the battery core. The thermal conductivity values may be further reduced upon cell aging.[57] Additional challenges are posed by the preference to heat or cool batteries quickly to limit preconditioning time as well as efforts to produce batteries with increased form factors which exhibit higher pack energy densities but have longer heat conduction paths.[58] Given that the temperature sensors employed by BTMSs are also external, it is not clear how sensitive current sensors are at identifying unsafe internal temperature variations. These challenges more broadly illustrate the inherent tradeoffs which must be made when optimizing for different battery metrics as well as the increasing importance of temperature uniformity which will become more challenging to realize for future batteries.

More broadly, the identification of lithium leaching from $LiC_6$ and $LiC_{12}$ in response to a temperature hotspot raises important safety concerns as the Li would not be detectable using conventional thermal management unless the hotspot itself overlapped with the regions of the battery probed by a temperature sensor. If the leached lithium was allowed to subsequently propagate during subsequent cycling as well as repeated hotspot exposure, it may be possible for the leached lithium to propagate until it causes a short which could also induce a positive feedback loop whereby the short generates heat which promotes Li deposition which generates more heat. Furthermore, the results of this study should expand understanding of the conditions conventionally thought to induce Li on the surface of graphite beyond Li plating observed during operation due to kinetic limitations at low temperatures and/or high C-rates. The related lithium leaching should be considered as our results show that it can be induced under relatively mild temperature heterogeneity.

Using *in situ* micro-Raman spectroscopy, *in situ* optical microscopy and COMSOL thermal simulations, we have identified thermally activated and localized lithium leaching from $LiC_6$ and $LiC_{12}$ upon inducing moderate temperature heterogeneity. The lithium leaching phenomenon appears to be dependent on the magnitude of temperature heterogeneity as well as the degree of lithiation. In order to explain our results, we have proposed a mechanism whereby heterogeneous degrees of lithiation are induced in response to temperature heterogeneity on the $LiC_6/LiC_{12}$ surface that can lead to lithium leaching. The results of this study have important implications for performance and especially safety considerations for future batteries as mild temperature heterogeneity in the absence of current has not to this point been thought of as a condition which could lead to Li formation on the surface of graphite via leaching.

**Associated Content**

**Supporting Information**

The Supporting Information is available free of charge at

Experimental details (materials used, electrode preparation, in situ cell fabrication, characterization methods), schematic of in situ cell, electrochemical data, additional optical microscopy images, and Raman spectra.

**Author Information**

**Corresponding Author**

**Yangying Zhu** – *Department of Mechanical Engineering, University of California, Santa Barbara, Santa Barbara, California 93106-5070, United States;*

https://orcid.org/0000-0001-91850316; Email: yangying@ucsb.edu

**Authors**

**Harrison Szeto** – *Department of Chemistry and Biochemistry, University of California, Santa Barbara, Santa Barbara, California 93106-5070, United States*

**Vijay Kumar** – *Department of Mechanical Engineering, University of California, Santa Barbara, Santa Barbara, California 93106-5070, United States*

**Author Contributions**

H. S. performed experiments, analyzed the results and prepared the manuscript. V. K. assisted in analysis. Y. Z. supervised, reviewed and edited the manuscript.

**Notes**

The authors declare no competing financial interest.

**Acknowledgements**

This work was supported by an Early Career Faculty grant from NASA's Space Technology Research Grants Program (grant number 80NSSC23K0072). The authors acknowledge the use of the Quantum Structures Facility within the California NanoSystems Institute, supported by the University of California, Santa Barbara and the University of California, Office of the President. The MRL Shared Experimental Facilities are supported by the MRSEC Program of the NSF under Award No. DMR 2308708; a member of the NSF-funded Materials Research Facilities Network.

Supporting Information:

*In situ* observation of thermally-activated and localized Li leaching from lithiated graphite

Harrison Szeto[1], Vijay Kumar[2], Yangying Zhu[2]

[1.] Department of Chemistry and Biochemistry, University of California Santa Barbara, Santa Barbara, CA 93106, USA
[2.] Department of Mechanical Engineering, University of California Santa Barbara, Santa Barbara CA 93106, USA

Email: yangying@ucsb.edu

**Experimental Section**
**Graphite electrode fabrication**
A graphite electrode was prepared by combining artificial graphite (MTI Lib-CMSG), polyvinylidene difluoride (PVDF) (Solef 5130), and carbon black (Timcal Super P) in a 90:5:5 weight ratio respectively in N-methyl-2-pyrrolidone (NMP) (Sigma Aldrich M79603). Specifically, PVDF was first dissolved in NMP and mixed in a speed mixer (Flacktek DAC 150.1 FV-K) for 30 minutes. Carbon black and graphite were added successively and mixed for 30 minutes each before being wet cast to 80 μm on 9 μm copper foil (MTI BCCF-9u) with a doctor blade (MTI SEKTQ50). The film was subsequently dried in a vacuum oven at 80 °C for a day before Ø12 mm disks were punched out using a disk cutter (MTI MSKT06). The graphite electrodes were subsequently transferred into an Ar glovebox for *in situ* cell fabrication.

**Cell fabrication**
An *in situ* Raman cell was prepared by punching out a 7/32" hole in the top cap of an SS304 2032 coin cell case (MTI Corporation, Item # CR2032CASE304) and epoxying (Loctite EA E-120HP) a Ø12 mm glass coverslip (Thorlabs CG15NH1) onto it. The glass coverslips were previously sonicated successively in acetone, isopropanol and deionized water for 15 minutes each before being dried with dry $N_2$ and placed in an $O_2$ plasma cleaner for 3 minutes. The epoxy was allowed to cure for a day before a Cu foil disk (16 mm diameter, 0.002" thick, 6 mm hole in center) was spot welded to the outer edge of the coin cell top case in order to establish electrical contact between the Li foil (to be placed upon it during cell assembly) and the coin cell case while maintaining optical access. The case was subsequently transferred into an Ar glovebox with $H_2O$ and $O_2$ levels below <0.3 ppm to fabricate an *in situ* Raman cell. The cell was fabricated in an inverted geometry in the following order, 2032 coin cell case top cap, 750 μm thick Li foil (Ø12 mm with Ø6 mm hole in center MTI EQ-Lib-LiF25), Ø19 mm separator with a 7/32" hole in the center (Celgard 2325), 90:5:5 graphite electrode, Ø15.5 mm × 0.06 mm thick stainless steel spacer (MTI CR20SPA05), stainless steel wave spring (CR20WS-SPR) and 2032 coin cell case bottom cap along with 500 μL of 1M $LiPF_6$ in 1:1 by vol. EC:DMC. The cell was rested overnight before being cycled.

**Cell cycling**
The *in situ* graphite half cell is cycled at C/10 on a potentiostat (Biologic SP-150). The rate of C/10 is chosen to establish robust EEI formation as well as ensure any Li identified on the surface of the graphite is attributed to the application of the laser hotspot and not any kinetically-limited Li plating at higher C-rates. The *in situ* cell is subsequently taken off at different stages during lithiation/delithiation and analyzed with *in situ* micro-Raman spectroscopy as well as optical microscopy.

**In situ micro-Raman spectroscopy**
*In situ* micro-Raman measurements were performed using a confocal micro-Raman spectrometer (Horiba Jobin Yvon T64000) equipped with a liquid nitrogen cooled CCD array detector (1024 × 256 pixels) and a single, 640 mm monochromator with a groove density of 1800 grooves/mm. An excitation wavelength of 488 nm is produced by an Ar/Kr ion laser (Coherent INNOVA 300C) that is subsequently focused on the sample surface with a 50x long working distance objective (Olympus LMPlanFL N50x). Backscattered light is subsequently transmitted through a high-pass filter and confocal slit set to 0.5 μm to limit band broadening. Spectra are acquired with an exposure time of 30 s and 30 accumulations; total acquisition time is 15 minutes over the spectral window. Spectra were baseline corrected and normalized relative to the most intense electrolyte peak (~ 894 cm$^{-1}$) in OriginLab.

Laser power densities are determined by measuring the laser power at the sample surface with a power meter (Thorlabs PM100D) equipped with a silicon (Si) photodiode (Thorlabs S130C) and estimating the laser spot size using the numerical aperture of the objective (NA = 0.5) and the wavelength of the excitation light (488 nm).

**COMSOL thermal simulations**

Thermal simulations utilizing the Heat Transfer in Solids module in COMSOL Multiphysics software were employed to estimate the temperature distribution induced by the application of the Raman laser. A 2D-axisymmetric model was produced which consists of a graphite particle embedded within a porous, composite graphite electrode and surrounded above by electrolyte. A heat transfer coefficient of h = 10 W/m$^2$K was applied as the natural convection boundary condition of the model. To simplify the model, the graphite particle is modelled based on the D50 particle size of the graphite used in the anode (r = 7.5 μm). The laser spot is modeled at the graphite-electrolyte interface as a constant power heat source on the surface of the graphite particle whose dimensions are determined based on the estimated laser spot size as described above. The temperature distribution is determined by solving the steady-state heat conduction equation (1) where Q is the laser power density and k is the thermal conductivity. The model is solved with a mesh size range of 0.1 to 1 μm.

$$Q = -k(\frac{\partial^2 T}{\partial x^2} + \frac{\partial^2 T}{\partial y^2} + \frac{\partial^2 T}{\partial z^2}) \quad (1)$$

Thermophysical properties for the electrolyte and graphite soaked in electrolyte at different stages of lithiation are taken from the literature and listed in Table S1.

**Table S1:** Thermophysical properties of components used in the COMSOL model.

|  | Heat capacity (J/kg·K) | Density (kg/m$^3$) | Thermal conductivity (W/m·K) |
|---|---|---|---|
| Graphite | 706.9[1] | 2260[2] | 77.52–216.63* |
| Electrolyte | 2055.1[3] | 1290[4]** | 0.2[5] |
| Composite graphite anode soaked w/ electrolyte | 1111[6] | 1907[6] | 0.96–1.29[7] |
| Copper | 385[8] | 8960[8] | 385[8] |
| Lithium | 4020[9] | 530[9] | 71.2[9] |
| Electrolyte-soaked separator | 1900[10] | 1009[10] | 0.5[10] |
| Stainless steel | 500[11] | 8000[11,12] | 16.2[11,12] |

\* Values estimated by converting thermal conductivity data for single crystal graphite at different degrees of lithiation to polycrystalline graphite. See Note S1 for more information on how the thermal conductivity values were obtained.

\*\*estimated from an electrolyte consisting of 1 M LiPF$_6$ in 1:1:1 EC:DMC:DEC

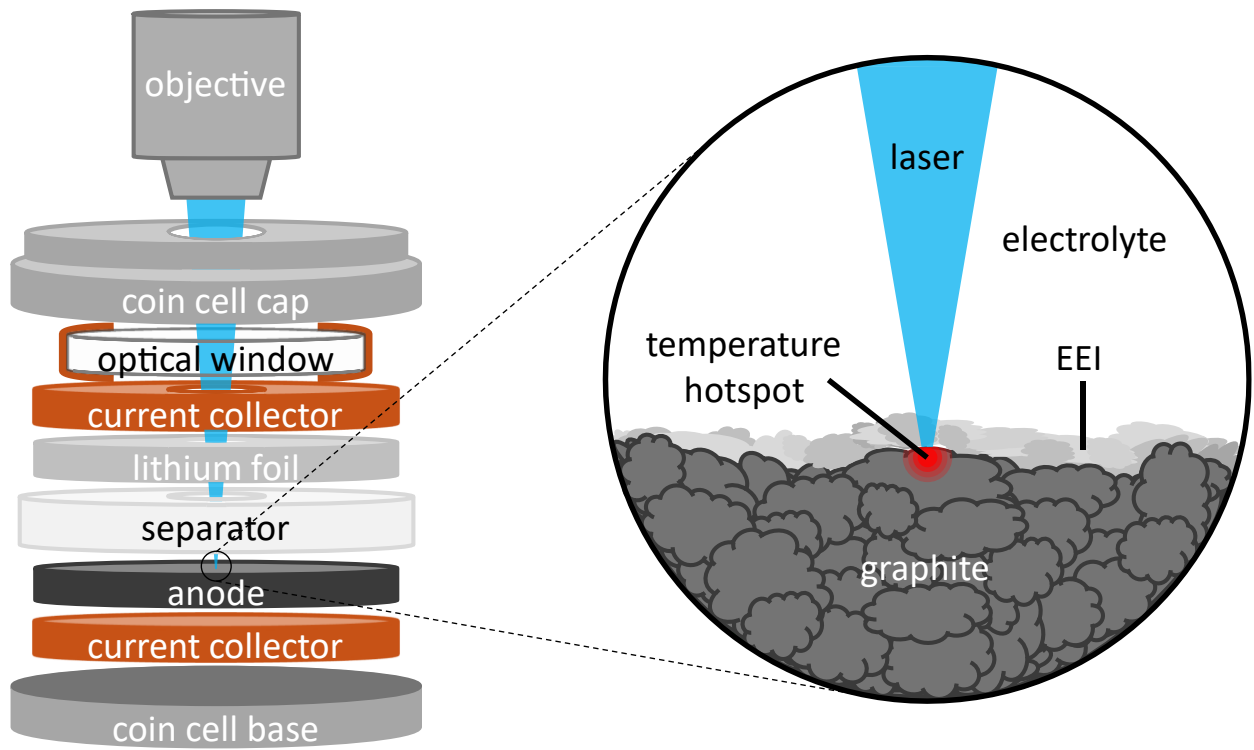

**Figure S1:** Diagram of in situ cell architecture with schematic of region probed using micro-Raman spectroscopy.

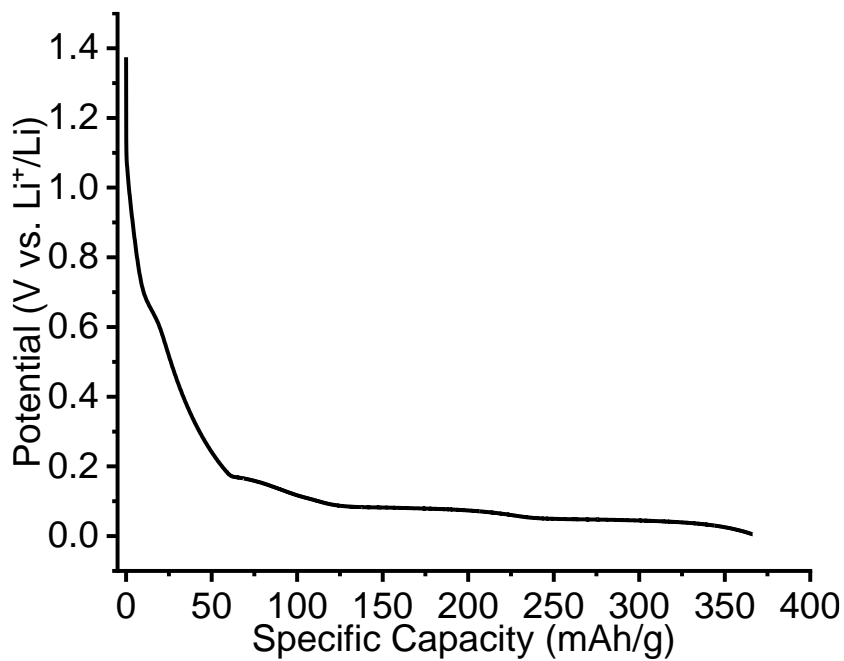

**Figure S2:** Electrochemical profile for initial C/10 lithiation of graphite.

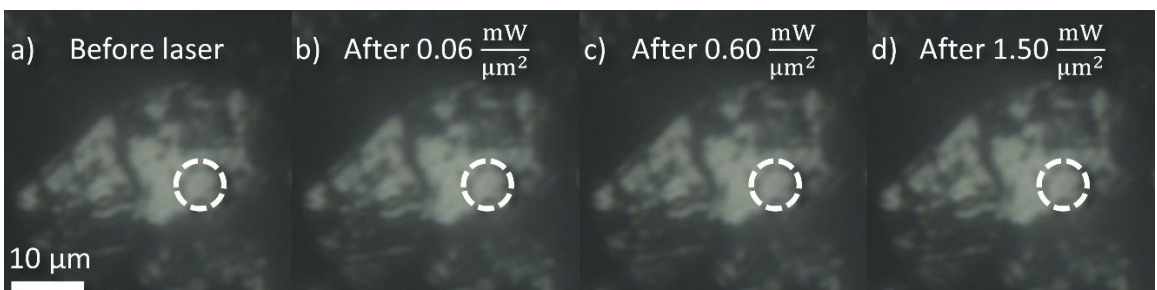

**Figure S3:** *in situ* optical microscopy images of unlithiated graphite (a) before and after laser exposure with varying power densities (b) 0.06 mW/μm², (c) 0.60 mW/μm², and (d) 1.50 mW/μm².

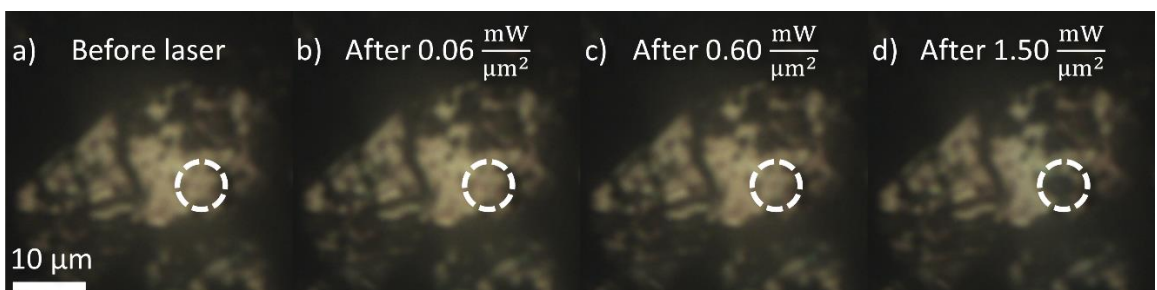

**Figure S4:** *in situ* optical microscopy images of fully lithiated graphite (LiC$_6$) (a) before and after laser exposure with varying power densities (b) 0.06 mW/μm², (c) 0.60 mW/μm², and (d) 1.50 mW/μm².

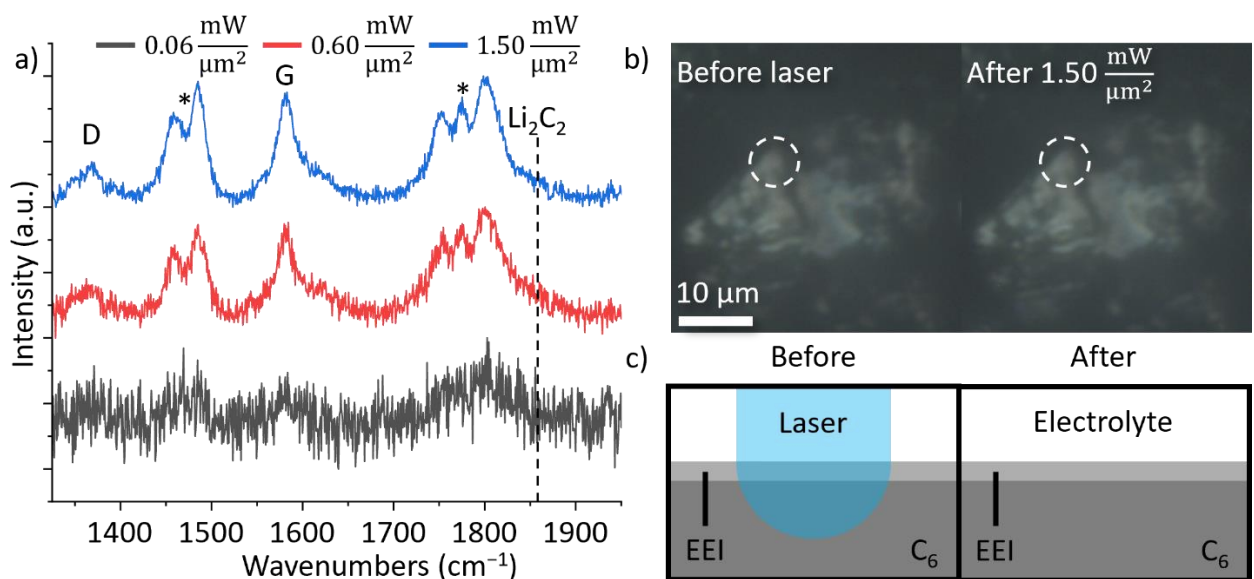

**Figure S5:** (a) Raman spectra for a graphite particle acquired following delithiation with different laser power densities. Note the lithium leaching evident in the neighboring region of the same graphite particle from prior exposure to the laser in the fully lithiated state. (b) Optical microscopy images taken before and after laser exposure for a delithiated graphite particle. Spectra taken in the delithiated state do not exhibit $Li_2C_2$ peak indicating that the species is not directly formed to an appreciable degree as part of the EEI of graphite and is therefore localized to the region exposed to the laser when lithiated. (c) Schematic depicting lithium leaching localized to the region of the graphite irradiated by the laser in the lithiated state.

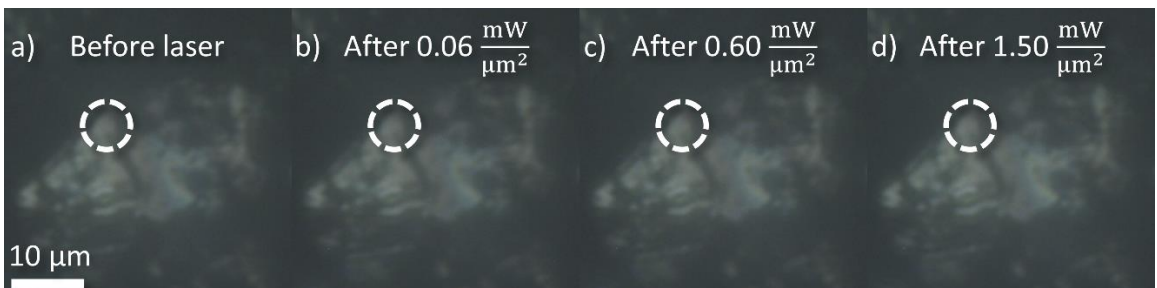

**Figure S6:** *in situ* optical microscopy images of delithiated graphite (a) before and after laser exposure with varying power densities (b) 0.06 mW/μm², (c) 0.60 mW/μm², and (d) 1.50 mW/μm².

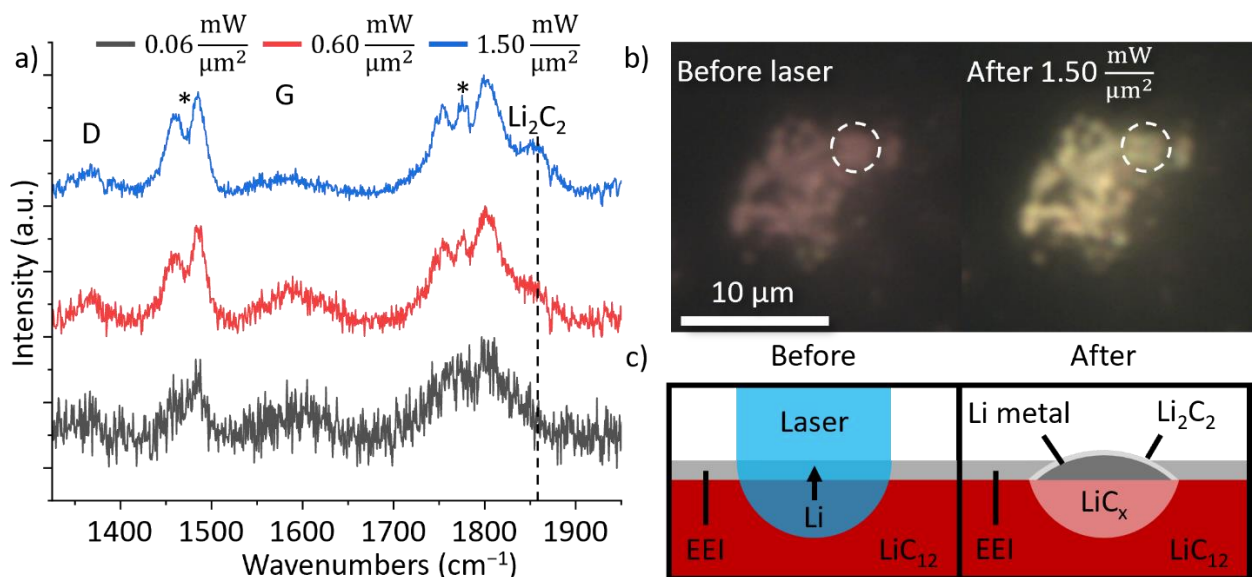

**Figure S7:** (a) Raman spectra for a graphite particle acquired in a partially lithiated state (approximately $LiC_{12}$) with different laser power densities. (b) Optical microscopy images taken before and after laser exposure for a partially lithiated (stage II) graphite particle. Note, that the red color indicates the particle is approximately in stage II as previously reported. (c) Schematic showing lithium leaching phenomenon from partially lithiated graphite.

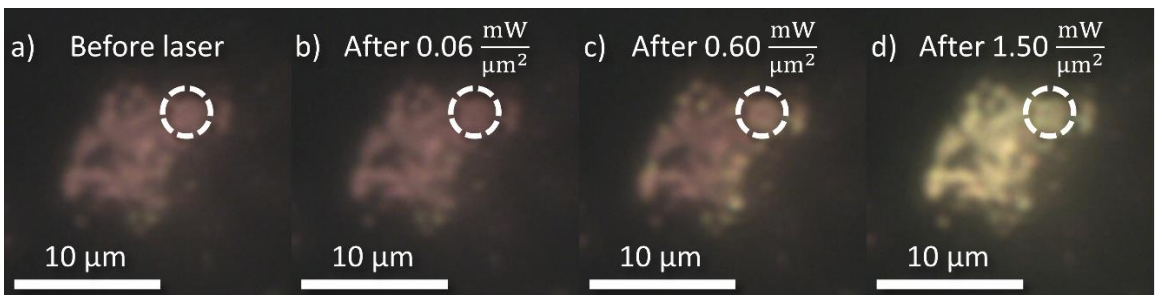

**Figure S8:** *in situ* optical microscopy images of fully lithiated graphite (LiC$_6$) (a) before and after laser exposure with varying power densities (b) 0.06 mW/μm$^2$, (c) 0.60 mW/μm$^2$, and (d) 1.50 mW/μm$^2$.

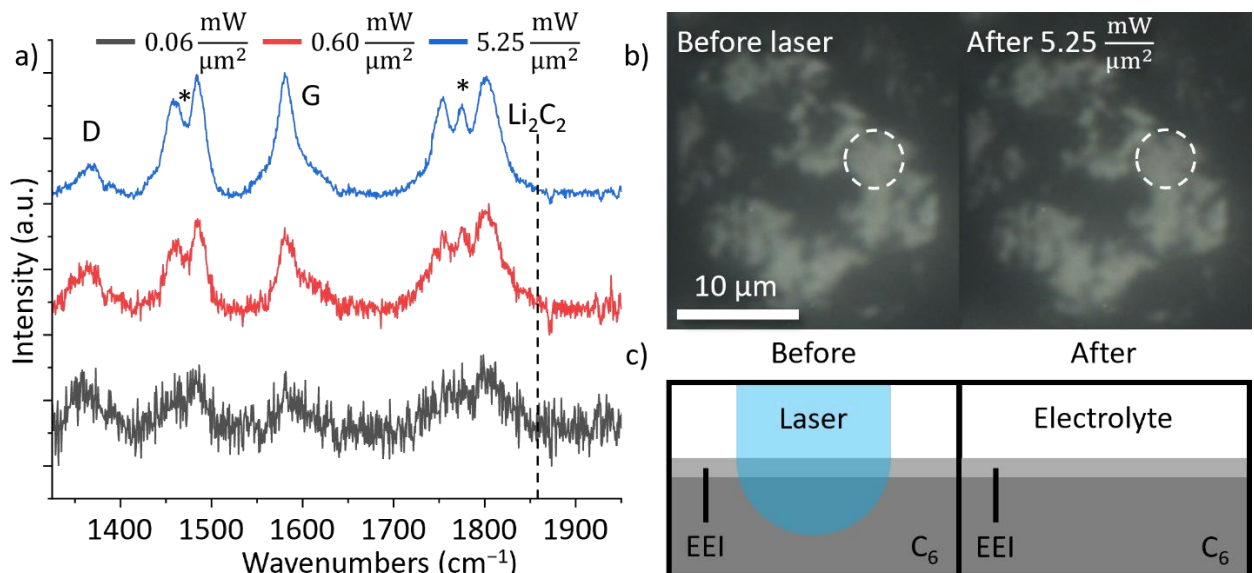

**Figure S9:** (a) Raman spectra of delithiated graphite under different laser power densities after uniform heating when lithiated. (b) optical microscopy images of a graphite particle after uniform heating and delithiation and before and after laser exposure (c) Schematic showing no lithium is leached from lithiated graphite when uniformly heating.

**Note S1:** Calculating the thermal conductivity of polycrystalline graphite at different degrees of lithiation from single crystal graphite data

During the lithiation of graphite, many of its properties change including thermal conductivity. We therefore use a method previously employed by Zeng et al in order to determine the thermal conductivity of the graphite particle during lithiation.[7] The method involves using the in-plane and through-plane thermal conductivities for graphite at different stages of lithiation that were determined from previous works employing molecular dynamics simulations. Previous works employing molecular dynamics simulations have found that Li intercalation results in an anisotropic tuning of single-crystal graphite's thermal conductivity.[13,14] While the Li intercalants were found to increase phonon scattering which reduces the in-plane thermal conductivity, the increased phonon scattering competes with an increase in phonon group velocity in the through-plane direction resulting in a non-monotonic change in the through-plane thermal conductivity. The in-plane and through-plane thermal conductivities of single-crystal graphite are subsequently converted into an effective thermal conductivity at different stages of lithiation using an equation proposed by Mityushov et al which assumes the graphite particle is polycrystalline with randomly oriented grains.[15] Similarly, the anode thermal conductivity which is a bulk property accounting for the graphite, binder, and conductive carbon has previously been measured and found to decrease during lithiation. These values are subsequently input into the COMSOL model to determine the temperature distribution induced by the micro-Raman laser at different stages of lithiation.

**Table S2:** List of Raman bands identified in spectra.

| Wavenumbers (cm$^{-1}$) | Heat capacity $C_p$ (J/kg·K) |
|---|---|
| ~1367 | Graphite D-band[16] |
| ~1459 | DMC[17] |
| ~1485 | EC[17] |
| ~1582 | G-band |
| ~1752 | DMC[17] |
| ~1773 | EC[17] |
| ~1800 | EC[17] |
| ~1857 | Li$_2$C$_2$[18] |